# Socioeconomics of Urban Travel in the U.S.: Evidence from the 2017 NHTS

Xize Wang[a, *], John L. Renne[b]

**Abstract**

Using the 2017 National Household Travel Survey (NHTS), this study analyzes America's urban travel trends compared with earlier nationwide travel surveys, and examines the variations in travel behaviors among a range of socioeconomic groups. The most noticeable trend for the 2017 NHTS is that although private automobiles continue to be the dominant travel mode in American cities, the share of car trips has slightly and steadily decreased since its peak in 2001. In contrast, the share of transit, non-motorized, and taxicab (including ride-hailing) trips has steadily increased. Besides this overall trend, there are important variations in travel behaviors across income, home ownership, ethnicity, gender, age, and life-cycle stages. Although the trends in transit development, shared mobility, e-commerce, and lifestyle changes offer optimism about American cities becoming more multimodal, policymakers should consider these differences in socioeconomic factors and try to provide more equitable access to sustainable mobility across different socioeconomic groups.

a. Department of Real Estate, National University of Singapore, Singapore. Email: wang.xize@nus.edu.sg. OCRID: 0000-0002-4861-6002
b. Department of Urban and Regional Planning, Florida Atlantic University, United States. Email: jrenne@fau.edu. OCRID: 0000-0002-1554-7557
* Corresponding author.

# 1 Introduction

This is the sixth article in a series of papers analyzing urban travel trends and their variations among a range of socioeconomic groups for the U.S., using nationwide travel survey data since 1977 (Pucher, Henderickson, McNeil, 1981; Pucher & Williams, 1992; Pucher, Evans and Wenger, 1998; Pucher & Renne, 2003; Renne and Bennett, 2014). In this article, we examine the 2017 National Household Travel Survey (NHTS), which was first released in 2018 and then updated in 2019 and 2020 (Federal Highway Administration, 2020; WeStat, 2019). We focus on the interrelated variations in travel modes, trip frequency, trip distance, and vehicle ownership, and their differences across different income levels as well as home ownership, ethnicity, gender, age, and life-cycle classifications. We also compare these variations in the 2017 NHTS with its predecessors, the Nationwide Personal Transportation Surveys (NPTS) in 1969, 1977, 1983, 1990, and 1995; and the National Household Travel Surveys (NHTS) in 2001 and 2009 based on earlier studies in this series (Pucher, Henderickson, McNeil, 1981; Pucher and Williams, 1992; Pucher, Evans and Wenger, 1998; Pucher and Renne, 2003; Renne and Bennett, 2014).

The most noticeable trend in Americans' travel behavior over the past decade is the peak and decline in the popularity of private automobile travel (Tables 1-2). According to journey-to-work data from the recent decennial census and the American Community Survey, the share of Americans taking private automobiles to work peaked in 2000 at 87.9%, then declined to 86.1% in 2009 and 85.3% in 2017 (see Table 1). The transit commuting share increased from 4.7% in 2000 to 5.1% in 2009 and remained at 5.0% in 2017. Additionally, the share of Americans working from home has also increased, before the COVID-19 pandemic, from 3.3% in 2000 to 4.3% in 2009 and 5.2% in 2017 (see Table 1). The trends in all-purpose travel (i.e., both work-and non-work-related) based on NPTS and NHTS data is similar (Table 2): the share of private automobile trips decreased from 86.4% in 2001 to 83.6% in 2009 and 82.6% in 2017. The share of transit trips increased from 1.6% to 2001 to 2.3% in 2009 and 2.9% in 2017.

**Table 1 Trends in Modal Split for the Journey to Work (1960–2017)**
**(percentage of work trips by mode of transportation)**

| Mode of Transportation | Census/ACS Year | | | | | | |
|---|---|---|---|---|---|---|---|
| | 1960 census | 1970 census | 1980 census | 1990 census | 2000 census | 2009 ACS 1yr | 2017 ACS 1yr |
| Total Auto | 66.9 | 77.7 | 84.1 | 86.5 | 87.9 | 86.1 | 85.3 |
|     SOV | n/a | n/a | 64.4 | 73.2 | 75.7 | 76.1 | 76.4 |
|     HOV | n/a | n/a | 19.7 | 13.4 | 12.2 | 10.0 | 8.9 |
| Public Transit | 12.6 | 8.9 | 6.4 | 5.3 | 4.7 | 5.1 | 5.0 |
| Walk | 10.3 | 7.4 | 5.6 | 3.9 | 2.9 | 2.9 | 2.7 |
| Bicycle | n/a | n/a | 0.5 | 0.4 | 0.4 | 0.6 | 0.5 |
| Work at Home | 7.5 | 3.5 | 2.3 | 3.0 | 3.3 | 4.3 | 5.2 |
| Other | 2.6 | 2.5 | 1.1 | 0.9 | 0.8 | 1.1 | 1.3 |
| All | 100 | 100 | 100 | 100 | 100 | 100.0 | 100 |

Sources: The 1960–2010 data are from the U.S. Census and drawn from (Renne & Bennett, 2014); the 2009 and 2017 data are from the American Community Survey one-year estimates and tabulated by the authors.
Notes: n/a = not available. To make the 1960 distributions comparable with those of later years, which do not include an "unreported" category, the 1960 reported modal shares were scaled up by a factor of 1.045 so that their total would equal approximately 100%.

**Table 2 Trends in Modal Split for Daily Travel in the United States**

| Mode of Transportation | 1969[a] | 1977 | 1983 | 1990 | 1995 | 2001 | 2009 | 2017 |
|---|---|---|---|---|---|---|---|---|
| Auto[b] | 81.8 | 83.7 | 82.0 | 87.1 | 86.5 | 86.4 | 83.6 | 82.6 |
| Transit | 3.2 | 2.6 | 2.2 | 2.0 | 1.8 | 1.6 | 2.3 | 2.9 |
| Walk[b] | n/a | 9.3 | 8.5 | 7.2 | 5.4 | 8.6 | 10.5 | 10.5 |
| Bicycle | n/a | 0.7 | 0.8 | 0.7 | 0.9 | 0.9 | 1.0 | 1.0 |
| Other[c] | 5.0 | 3.7 | 6.5 | 3 | 5.4 | 2.5 | 2.6 | 3.0 |

Sources: Federal Highway Administration, Nationwide Personal Transportation Surveys 1969, 1977, 1983, 1990, and 1995; National Household Travel Surveys 2001, 2009, and 2017; Pucher, Henderickson and McNeil (1981); Pucher and Williams (1992); Pucher, Evans and Wenger (1998); Pucher and Renne (2003); and Renne and Bennett, (2014)

Notes: n/a = not available. Unlike all subsequent tables, these NPTS and NHTS modal split percentages are for daily, local travel in aggregate for the entire USA, both urban and rural, as reported by the FHWA in its own NPTS and NHTS reports. Our own tabulations, from Table 3 onward, include only local trips in urban areas.
a. The 1969 NPTS did not sample walk and bike trips, thus artificially inflating the modal split shares of the motorized modes compared to the NPTS surveys in later years. To ensure some degree of comparability, we adjusted downward the reported motorized shares of trips in 1969 by 10%, using the percentage of walk and bike trips in 1977. That is why the column adds up to 90% and not 100%. Our adjustment is rough, but otherwise, the 1969 and later NPTS modal split distributions would be completely incomparable.
b. The decrease in auto mode share from 1995 to 2001, and the corresponding increase in walk mode share during the same period, are due to a change in sampling methodology that captures previously unreported walk trips.

Such trends in private automobile travel are reflected by changes in automobile ownership. The number of vehicles per licensed driver peaked in 2001 at 1.06, then dropped to 0.99 in 2009 and remained at 1.00 in 2017. Similarly, vehicles per household peaked in 2001 at 1.89 and then dropped to 1.86 in 2009, although bounced back to 1.88 in 2017. Notably, the average household size also increased from 2.50 in 2009 to 2.55 in 2017 (McGuckin and Fucci, 2018; Pucher and Renne, 2003; Renne and Bennett, 2014). In other words, although the American economy has been emerging from the 2008 recession, private automobile ownership in 2017 did not bounce back to pre-recession levels. This trend is even more apparent among urban dwellers, who constitute an increasing share of the nation's total population during the first two decades of the 21st century (United Nations, 2018). The percentage of urban households owning at least one automobile in 2017 was 91.3%, lower than the peak value in 2001 (91.7%); the rate of urban households owning more than one automobile in 2017 wass 56.0%, which was 2.5 percentage points lower than the historical peak value in 2001 (58.5%).

Such trends can be attributed to many factors, including the expansion of transit-oriented developments (TOD) in urban America (Renne and Appleyard, 2019; Boarnet, Wang and Houston, 2017; Ewing and Cervero, 2010), intergenerational lifestyle changes (Garikapati et al., 2016; McDonald, 2015; Blumenberg et al, 2019), the increase of fuel price (Bastian, Borjesson and Eliasson, 2016; Stapleton, Sorrell and Schwanen, 2017), the rise of e-commerce and shared mobility (Brown, 2019; Cao, 2012; Dong, 2020; Le, Carrel and Shah, 2022), and increasing technologies that have enabled remote working (Aksoy et al., 2022; Su, McBride and Goulias, 2021). With the continuation of many of these factors, and recent changes due to the pandemic and the "Great Resignation," such as flexible work arrangements, it is an open and interesting question whether this trend in American urban travel will continue.

Nevertheless, such an aggregate trend may mask complex variations in travel patterns among different social and demographic sub-groups. Theoretical reasoning and empirical evidence support the fact that travel behavior differs by social and demographic profiles such as income, age, gender and ethnicity (Ben-Akiva and Lerman, 1985; de Dios Ortúzar and Willumsen, 2011; Pucher and Renne; 2003; Renne and Bennett, 2014). Vehicle ownership levels, trip frequencies, miles traveled and means of transport vary from one group to another, which is increasingly becoming a vital data issue for examining equity in travel. Examining such differences can help policy makers to better evaluate the effects of existing transportation infrastructure developments, to identify the sub-groups that deserves special attention for future transport plans, and to design more sustainable and equitable transport policies at local, state and federal levels. The following sections of this article introduces the 2017 NHTS, examines variations in urban travel with respect to various socioeconomic factors, compares with earlier surveys and discusses the policy implications.

## 2    Data and methods

The 2017 NHTS is the most recent nationally representative travel survey in the United States. The data collection for this survey started in April 2016 and ended in April 2017 (Westat, 2019). The survey used an address-based sample and collected data from approximately 130,000 households and 275,000 persons; the study sample is nationally representative, although a few states and metropolitan areas were oversampled since the respective transportation planning agencies had purchased "add-on" samples (McGuckin and Fucci, 2018; Westat, 2019). However, the NHTS provides weights for each household and each person in the study sample to adjust for such oversampling (Westat, 2019), and this article uses the respective weights to make the study sample nationally representative. Each participating household reported all trips by household members on a randomly assigned 24-hour "travel day," which could be either a weekday or a weekend day. The respondents then used a web-based survey interface to report their travel history on that "travel day," as well as other information, such as socio-demographic details, economic conditions, place of residence, and more (Westat, 2018). The full list of variables in the 2017 NHTS public sample is available at the NHTS website developed by the Oak Ridge National Laboratory (ORNL, 2022).

The analyses in this article are based on the travel behavior data reported in the "travel day" and adjusted for household-, personal- and trip-level weights provided by the 2017 NHTS. Since this study focuses on urban travel, the study sample is limited to residents in urban areas. Trips longer than 75 miles were excluded, which is consistent with the previous studies in this series. The resulting sample includes 100,465 households and 202,907 individuals living in the urban area, as well as 712,198 urban trips. We start our analysis with examining the differences of urban travel with respect to trip purpose (e.g., work- and non-work-related) and geographical regions (e.g., New England, Pacific, and others); then we study the effect of economic factors (income, vehicle ownership, housing) on urban behaviors in the cities; finally, we examine the variations in urban travel by demographic factors (gender, ethnicity, age and life-cycle stage).

There are a few differences in the 2017 NHTS compared with earlier surveys (Westat, 2019). First, the 2017 NHTS differentiates "round trips" from "one-way" trips, while earlier surveys only recognized one-way trips. Thus, we split each round trips into two trips (and halved the distance) to make it comparable with earlier surveys. Second, the 2017 NHTS used an online mapping tool to estimate trip distances, which was different from the previous NHTS. To make the trip distance variables comparable with earlier surveys, the 2017 NHTS reports both the unadjusted (online mapping-tool derived) and adjusted (inflated by roughly 10% to be comparable with earlier NHTS) trip distances. In this article, we report both distances but mainly discuss the unadjusted ones. Additionally, a few categorical variables in the 2017 NHTS are defined differently from in earlier surveys, such as household income, trip purpose, and travel modes; we try our best to combine different categories in our analyses to make this study comparable with those based on earlier surveys, although a few variables still are not comparable across the different surveys. For detailed discussions on comparing the categorical variables of the 2017 NHTS with earlier NHTS/NPTS, see Westat (2019).

## 3    Results

### 3.1    *Trip purpose and modal choice*

As previously mentioned, the share of work- and non-work-related transit trips in the United States has steadily increased over the first 17 years of the 21st century. In 2017, transit trips comprise 5% of the total commute trips (see Table 1), it is larger than the share of transit trips of total trips (2.9%, Table 2). Similarly, the share of transit trips for work-related travel is the highest, at 5.9%; specifically, the shares of bus trips and non-commuter-rail transit trips (i.e., subway, light rail, and streetcar) are 2.7% and 2.5%, respectively (Table 3). In other words, the shares of bus trips and subway/light rail/streetcar trips for work-related purposes are comparable. In contrast, most of the transit trips for shopping and services, social and recreational activities, and school and church purposes were made by bus. Additionally, 0.7% of total work-related trips were made by commuter rail, and the share of commuter rail trips for other purposes were much lower (0.1–0.2%, Table 3). Thus, the percentage of rail transit, including commuter rail, is highest for work-related trips.

Having that said, 82.6% of work-related trips were still made in private automobiles; solo drivers making up 66.3% of these and only 16.3% by carpools (see Table 3). In contrast, carpooling was the predominant mode for trips for the other three purposes (i.e. shopping and services, social and recreation, and school and church), which made up almost half of all trips. These carpooling trips for shopping and services, social and recreational activities, and school and church were typically made by family. In contrast, the low share of carpooling trips for work-related purposes implies that most family members could not carpool or chose not to carpool in their journeys to work. The percentage of non-motorized trips was the highest for social and recreational trips; almost one in every four of such trips (23.4%) was made by walking or cycling. This is followed by school and church trips, 12.4% of which were non-motorized. The share of walking and cycling trips was lower for work- and shopping-related trips, at around 10%.

**Table 3 Variations in Modal Choice by Trip Purpose**

| Mode of Transportation | Trip Purpose | | | |
|---|---|---|---|---|
| | Work and Work-Related | Shopping and Services | Social and Recreation | School and Church |
| Total Auto | 82.6 | 87.5 | 72.1 | 69.2 |
|     SOV[a] | 66.3 | 40.8 | 25.2 | 17.5 |
|     HOV[b] | 16.3 | 46.8 | 46.9 | 51.7 |
| Total Transit | 5.9 | 2.4 | 2.8 | 3.6 |
|     Bus Transit[c] | 2.7 | 1.8 | 1.7 | 2.6 |
|     Subway/Light Rail/Streetcar[d] | 2.5 | 0.5 | 0.9 | 0.9 |
|     Commuter Rail[e] | 0.7 | 0.1 | 0.2 | 0.1 |
| Total Non-motorized | 9.4 | 9.7 | 23.4 | 12.4 |
|     Walk | 8.3 | 9.0 | 21.3 | 11.4 |
|     Bicycle | 1.1 | 0.7 | 2.1 | 1.1 |
| School Bus | 0.2 | 0.0 | 0.3 | 14.3 |
| Taxicab | 0.8 | 0.2 | 0.8 | 0.2 |
| Other | 1.1 | 0.3 | 0.6 | 0.2 |
| All | 100.0 | 100.0 | 100.0 | 100.0 |

Source: Calculated by the authors from the 2017 NHTS.
Notes: In order to isolate urban travel, the sample was limited to residents of urban areas and trips of 75 miles or less.
a. SOV (single-occupancy vehicle) includes vehicles with driver and no passengers.
b. HOV (high-occupancy vehicle) includes vehicles with two or more occupants.
c. Bus transit includes public bus, commuter bus, private/shuttle bus, and city-to-city bus.
d. Subway/light rail/streetcar also includes elevated rail.
e. Commuter rail includes suburban/regional rail systems and short-distance service provided by Amtrak.

*3.2 Regional variations in transit use, walking, and cycling*

The shares of transit and non-motorized trips also varied by region, which is shaped by the unique historical, cultural and urban design characteristics of each region of the United States. Here we report the regional variations in transit and non-motorized modal shares at the largest scale (i.e., census regions, Table 4); due to space limits, we report the MSA- and state-level analyses in the Appendix (Tables A1 and A2). As Table 4 shows, the Middle Atlantic region, where New York City and Philadelphia are located, sees the largest share of transit trips (8.5%). It is followed by two highly urbanized regions — New England and the Pacific — with shares of 4.7% and 3.3%, respectively. These three regions also have the highest shares of non-motorized trips, with 22.8% for Middle Atlantic, 16.6% for New England, and 15.3% for Pacific. Most of the non-motorized trips are walking trips, with the share as high as 21.8% for Middle Atlantic. The shares of bicycle trips are the highest in the Mountain (1.8%), Pacific (1.4%), and West North Central (1.4%) regions. In contrast, the East South Central region has the lowest share of transit and non-motorized trips (1.3% and 8.4%, respectively), as this region has relatively fewer transit-oriented or non-motorized friendly cities (Bureau of Transportation Statistics, 2022).

**Table 4 Regional Variations in Modal Shares for Transit, Walking, and Bicycling (percentage of trips by mode)**

| Mode of Transportation | New England | Middle Atlantic | East North Central | West North Central | South Atlantic | East South Central | West South Central | Mountain | Pacific |
|---|---|---|---|---|---|---|---|---|---|
| Total Transit | 4.7 | 8.5 | 3.2 | 1.6 | 2.3 | 1.3 | 1.4 | 1.8 | 3.3 |
|     Bus Transit | 2.5 | 3.4 | 2.1 | 1.4 | 1.7 | 1.3 | 1.3 | 1.5 | 2.4 |
|     Subway/Light Rail/Streetcar | 1.7 | 4.4 | 0.8 | 0.2 | 0.6 | 0.0 | 0.2 | 0.2 | 0.7 |
|     Commuter Rail | 0.4 | 0.8 | 0.3 | 0.0 | 0.1 | 0.0 | 0.0 | 0.0 | 0.2 |
| Total Non-motorized | 18.0 | 22.8 | 13.2 | 11.1 | 12.3 | 8.4 | 9.9 | 13.6 | 15.3 |
|     Walk | 16.6 | 21.8 | 12.2 | 9.7 | 11.4 | 8.0 | 9.0 | 11.9 | 13.8 |
|     Bicycle | 1.3 | 1.1 | 1.0 | 1.4 | 0.9 | 0.4 | 0.9 | 1.8 | 1.4 |

Source: Calculated by the authors from the 2017 NHTS.
Note: In order to isolate urban travel, the sample was limited to residents of urban areas and trips of 75 miles or less. Region is defined as census regions by the U.S. Census Bureau, for specific definitions, see:
https://www2.census.gov/geo/pdfs/maps-data/maps/reference/us_regdiv.pdf

## 3.3 Economic factors and urban travel

### 3.3.1 Income, vehicle ownership and mobility levels

This subsection examines the variation in mobility (trip frequency and distances) and vehicle ownership by income levels. Table 5 shows the average daily trip frequencies and distances by household income groups. For trip distance measures, as mentioned in Section 2, the 2017 NHTS reports both unadjusted and adjusted trip distances, and the adjusted distances are aimed at being comparable with earlier NHTS. Here we report both measures, and use the unadjusted trip distances in our analyses. As indicated in Table 5, although the number of trips per day does not vary much across household income levels, miles traveled per person varies by household income. Specifically, the average miles traveled per day for the lowest income group (less than $25,000) was 36% lower (14.3 miles vs. 19.5 miles) than the second-lowest income group ($25,000–$49,999). Conversely, although miles traveled per person was higher for higher-income groups, the differences among other higher income categories were smaller than that between the lowest vs. second-lowest income groups. The average daily trip distances of all other higher income groups (i.e. $50,000 and higher) were between 21.8 miles and 24.6 miles.

**Table 5 Daily Travel per Capita by Income Class**

| Household Income (per year) | Trips per Day, per Person | Miles Traveled (unadjusted) per Day, per Person | Miles Traveled (adjusted) per Day, per Person |
|---|---|---|---|
| Less than $25,000 | 3.3 | 14.3 | 15.6 |
| $25,000 to $49,999 | 3.6 | 19.5 | 21.3 |
| $50,000 to $74,999 | 3.6 | 21.8 | 23.9 |
| $75,000 to $99,999 | 3.7 | 23.7 | 25.9 |
| $100,000 and over | 3.8 | 24.6 | 26.9 |
| All | 3.6 | 20.9 | 22.9 |

Source: Calculated by the authors from the 2017 NHTS.
Note: In order to isolate urban travel, the sample was limited to residents of urban areas and trips of 75 miles or less.

Table 6 reports the average trip length by mode and income class. Due to space limits, here we only report the unadjusted trip distances; the same table using adjusted trip distances is available in the Appendix (see Table A3). As Table 6 indicates, those with higher income generally took longer trips. Similar to the above-mentioned daily miles traveled, the lowest income group had a considerably lower average trip distance (4.6 miles) than the second-lowest income group (5.7 miles), and the other higher income groups' average trip distances were greater than 6 miles (6.3–6.7 miles). Additionally, the average trip length for the highest income group (6.7 miles) is 1.46 times of that of the lowest income group (4.6 miles). Such high–low income ratio in trip distance is smaller for automobile trips: the average trip distance of the highest income group (7.5 miles) is 1.36 times that of the lowest income group (5.5 miles). In contrast, this high–low income ratio in trip distance is larger for transit trips: the average trip distance for the highest income group (12.8 miles) is 1.91 times that of the lowest income group (6.7 miles). Specifically, although the average distances of subway/light rail/streetcar trips are similar across different income categories, the high–low income ratio in trip distances is large in bus transit (11.9 miles vs. 6.1 miles, including commuter buses) and commuter rail (25.5 miles vs. 18.0 miles). Such facts imply that many of the transit trips by the highest income group are commute trips on buses and rail. Commuter buses and trains are almost exclusively used by suburban residents to commute to and from central city destinations, which helps to reduce peak-hour congestion and carbon emissions but may offer limited help in improving accessibility for inner-city lower income residents.

**Table 6 Average Trip Length (unadjusted) by Mode and Income Class (in miles)**

| Mode of Transportation | Household Income | | | | | |
|---|---|---|---|---|---|---|
| | Less than $25,000 | $25,000 to $49,999 | $50,000 to $74,999 | $75,000 to $99,999 | $100,000 and over | All |
| Total Auto | 5.5 | 6.4 | 7.1 | 7.6 | 7.5 | 6.9 |
|     SOV[a] | 5.6 | 6.3 | 7.3 | 7.9 | 8.1 | 7.2 |
|     HOV[b] | 5.5 | 6.5 | 6.9 | 7.3 | 7.0 | 6.7 |
| Total Transit | 6.7 | 8.2 | 9.2 | 9.7 | 12.8 | 9.1 |
|     Bus Transit[c] | 6.1 | 6.8 | 7.9 | 9.7 | 11.9 | 7.5 |
|     Subway/Light Rail/Streetcar[d] | 8.6 | 10.2 | 9.6 | 8.2 | 9.3 | 9.2 |
|     Commuter Rail[e] | 18.0 | 17.0 | 21.1 | 23.6 | 25.5 | 23.4 |
| Total Non-motorized | 0.8 | 0.7 | 0.7 | 0.7 | 0.7 | 0.7 |
|     Walk | 0.7 | 0.6 | 0.5 | 0.5 | 0.6 | 0.6 |
|     Bicycle | 1.6 | 2.4 | 2.1 | 2.2 | 2.3 | 2.1 |
| School Bus | 3.7 | 3.8 | 4.6 | 3.8 | 4.4 | 4.1 |
| Taxicab | 4.8 | 4.6 | 3.7 | 7.8 | 6.8 | 5.7 |
| Other | 3.9 | 10.5 | 9.1 | 7.1 | 10.7 | 8.3 |
| All | 4.6 | 5.7 | 6.3 | 6.7 | 6.7 | 6.1 |

Source: Calculated by the authors from the 2017 NHTS.
Notes: In order to isolate urban travel, the sample was limited to residents of urban areas and trips of 75 miles or less.
a. SOV (single-occupancy vehicle) includes vehicles with driver and no passengers.
b. HOV (high-occupancy vehicle) includes vehicles with two or more occupants.
c. Bus transit includes public bus, commuter bus, private/shuttle bus, and city-to-city bus.
d. Subway/light rail/streetcar also includes elevated rail.
e. Commuter rail includes suburban/regional rail systems and short-distance service provided by Amtrak.

As indicated above, the largest gaps in daily miles traveled and average trip distance occur between the lowest and the second-lowest income groups. Table 7 illustrates a similar trend in automobile ownership as 29% of the lowest-income households (less than $25,000) did not own any vehicles, compared with only 6.1% for the second-lowest income group, and no more than 4% for other higher income groups. Many households still need at least one private automobile to meet their daily mobility needs (Brown, 2017; King, Smart and Manville, 2022). Additionally, the share of households owning more than one vehicle increases monotonically with household income, with the share of the lowest income group at 20.4%, that of the second-

lowest income group at 43.5%, and the shares of the other three income groups at 60.4%, 70.3%, and 81.5%, respectively. The share of no-vehicle and one-vehicle households increased from 2009 to 2017, while the share of two or more vehicles per household decreased from 2009 to 2017. This trend of declining vehicle ownership despite the economic recovery may be attributed to many factors, including the rapid growth of transportation network companies (TNCs) and micromobility services, the wider adoption of travel demand management such as congestion pricing schemes, the surge in online shopping and generational differences in lifestyles leading urban residents to prefer owning fewer vehicles (Brown, 2019; Dong, 2020; Le, Carrel and Shah, 2022; Lime, 2019; Manville, 2021; Smart and Klein, 2018; Wang, 2019).

**Table 7 Vehicle Ownership by Income Class (percentage distribution within each income class)**

| Vehicles per Household | Household Income | | | | | |
|---|---|---|---|---|---|---|
| | Less than $$25,000 | $25,000 to $49,999 | $50,000 to $74,999 | $75,000 to $99,999 | $100,000 and over | All |
| 0 | 29.0 | 6.1 | 4.0 | 3.3 | 2.7 | 9.9 |
| 1 | 50.5 | 50.5 | 35.6 | 26.4 | 15.8 | 36.1 |
| 2 | 14.5 | 29.4 | 38.6 | 43.5 | 44.9 | 33.1 |
| 3 or more | 5.9 | 14.1 | 21.8 | 26.8 | 36.6 | 20.8 |
| Total | 100 | 100 | 100 | 100 | 100 | 100 |

Source: Calculated by the authors from the 2017 NHTS.
Note: In order to isolate urban travel, the sample was limited to residents of urban areas.

Table 8 looks at the mode shares of different vehicle ownership levels. Unsurprisingly, the biggest difference in private automobile trip shares is between those from no-vehicle households and those from one-vehicle households. For those from no-vehicle households, only 19.9% of the total trips were by automobiles, and a vast majority of these trips (17.1%) were by carpooling. Most of the people in these carpooling trips were likely to be carpool passengers. Only 2.9% of the total trips for this group is by solo-driving, which are mostly through borrowing from friends and relatives, car rentals or car sharing programs (Iacobucci, 2022; Lovejoy & Handy, 2011; Martin, Shaheen & Lidicker, 2010). For those from one-vehicle households, 76.9% of the trips were by private automobiles, and the share of carpooling trips (41.5%) was slightly higher than that of solo driving (35.3%). The shares of private automobile trips for two-car and three-or-more-car households was 84.8% and 88.1%, respectively. Similarly, the largest differences in transit trip shares also occur between zero-vehicle and one-vehicle households. For those from a zero-vehicle household, one in every four trips was by transit; the share of transit trips for one-vehicle households was only 3.7%. This fact shows that for many one-vehicle households, one person is the primary car user, and other household members need to take transit to fulfill their mobility needs. Finally, 47.9% of the trips made by no-vehicle households were non-motorized — most were walking trips; in comparison, the share of non-motorized trips for one-vehicle households was only 16.7%.

**Table 8 Impact of Auto Ownership on Mode Choice (percentage of trips by means of transportation)**

| Mode of Transportation | Total Number of Vehicles in Household | | | | |
|---|---|---|---|---|---|
| | 0 | 1 | 2 | 3 or more | All |
| Total Auto | 19.9 | 76.9 | 84.8 | 88.1 | 79.7 |
|     SOV[a] | 2.9 | 35.3 | 37.7 | 45.3 | 37.1 |
|     HOV[b] | 17.1 | 41.5 | 47.1 | 42.8 | 42.6 |
| Total Transit | 25.4 | 3.7 | 1.4 | 1.0 | 3.3 |
|     Bus Transit[c] | 17.1 | 2.1 | 0.7 | 0.5 | 2.0 |
|     Subway/Light Rail/Streetcar[d] | 7.7 | 1.4 | 0.4 | 0.3 | 1.1 |
|     Commuter Rail[e] | 0.7 | 0.2 | 0.2 | 0.2 | 0.2 |
| Total Non-motorized | 47.9 | 16.7 | 11.3 | 8.7 | 14.2 |
|     Walk | 44.5 | 15.3 | 10.4 | 8.0 | 13.1 |
|     Bicycle | 3.4 | 1.4 | 0.9 | 0.7 | 1.1 |
| School Bus | 2.6 | 1.6 | 1.8 | 1.6 | 1.7 |
| Taxicab | 2.9 | 0.6 | 0.4 | 0.3 | 0.6 |
| Other | 1.3 | 0.5 | 0.3 | 0.5 | 0.4 |
| All | 100 | 100 | 100 | 100 | 100 |

Source: Calculated by the authors from the 2017 NHTS.

Notes: In order to isolate urban travel, the sample was limited to residents of urban areas and trips of 75 miles or less.
a. SOV (single-occupancy vehicle) includes vehicles with driver and no passengers.
b. HOV (high-occupancy vehicle) includes vehicles with two or more occupants.
c. Bus transit includes public bus, commuter bus, private/shuttle bus, and city-to-city bus.
d. Subway/light rail/streetcar also includes elevated rail.
e. Commuter rail includes suburban/regional rail systems and short-distance service provided by Amtrak.

3.3.2 Income and travel modes

This subsection explores the impact of income on travel mode choice, as well as the variations of transit mode share across different city sizes. Table 9 shows the variations in mode shares by different income groups. Such variation is the composite of income effects and vehicle ownership effects. The biggest difference in automobile mode share comes from the lowest income group compared to the second-lowest income group: for the income group "less than $25,000/year," 69.6% of the total trips were made by automobiles, whereas for the "$25,000–$49,000/year" group, the share of auto trips was 82.6%. Indeed, except for the lowest income group, all other income groups made at least 80% of their trips in private automobiles. Specifically, the share of carpooling trips across these different income categories was very similar (40.4~43.9%), and the differences in automobile mode shares are mostly from variations in the shares of solo driving trips.

Such a "lowest income group vs. the others" disparity also applies to other trip modes. For instance, as Table 9 shows, the share of transit trips was 6.8% for the lowest income group and 2.4–2.6% for other income groups. Specifically, the share of bus transit trips was lower for higher-income households: 5.7% of the total trips for the lowest-income group were by bus, while that share was 1.7% or lower for other groups. In contrast, the highest income group ($100,000 and above) had the highest share of subway/light rail/streetcar and commuter trips (1.7% total). In other words, a typical bus user profile is someone from a low-income household

who does not own a vehicle; and a typical rail user profile is someone from a relatively higher-income household.

**Table 9 Modal Split by Income Class (percentage of trips by means of transportation)**

| Mode of Transportation | Household Income | | | | | |
| --- | --- | --- | --- | --- | --- | --- |
| | Less than $25,000 | $25,000 to $49,999 | $50,000 to $74,999 | $75,000 to $99,999 | $100,000 and over | All |
| Total Auto | 69.6 | 82.6 | 82.5 | 83.0 | 80.8 | 79.7 |
|     SOV[a] | 29.2 | 39.9 | 40.4 | 39.1 | 37.2 | 37.1 |
|     HOV[b] | 40.4 | 42.7 | 42.0 | 43.9 | 43.6 | 42.6 |
| Total Transit | 6.8 | 2.6 | 2.4 | 2.4 | 2.6 | 3.3 |
|     Bus Transit[c] | 5.7 | 1.7 | 1.2 | 1.2 | 0.9 | 2.0 |
|     Subway/Light Rail/Streetcar[d] | 0.9 | 0.8 | 1.1 | 1.1 | 1.3 | 1.1 |
|     Commuter Rail[e] | 0.2 | 0.1 | 0.1 | 0.1 | 0.4 | 0.2 |
| Total Non-motorized | 19.7 | 12.5 | 12.4 | 12.2 | 14.1 | 14.2 |
|     Walk | 18.1 | 11.6 | 11.4 | 11.2 | 12.9 | 13.1 |
|     Bicycle | 1.5 | 0.9 | 1.1 | 1.1 | 1.2 | 1.1 |
| School Bus | 2.7 | 1.5 | 1.5 | 1.6 | 1.5 | 1.7 |
| Taxicab | 0.6 | 0.3 | 0.6 | 0.5 | 0.7 | 0.6 |
| Other | 0.7 | 0.5 | 0.5 | 0.2 | 0.3 | 0.4 |
| All | 100.0 | 100.0 | 100.0 | 100.0 | 100.0 | 100 |

Source: Calculated by the authors from the 2017 NHTS.
Notes: In order to isolate urban travel, the sample was limited to residents of urban areas and trips of 75 miles or less.
a. SOV (single-occupancy vehicle) includes vehicles with driver and no passengers.
b. HOV (high-occupancy vehicle) includes vehicles with two or more occupants.
c. Bus transit includes public bus, commuter bus, private/shuttle bus, and city-to-city bus.
d. Subway/light rail/streetcar also includes elevated rail.
e. Commuter rail includes suburban/regional rail systems and short-distance service provided by Amtrak.

Table 10 provides a closer look at urban transit users by metropolitan area size and income levels. First, metropolitan areas with populations of three million or more were the only group that consistently had higher income-group-specific transit mode shares than the national averages across all income categories. For instance, the nationwide transit mode shares for the lowest and highest income group are 6.8% and 2.6%, respectively; for the largest metropolitan areas (three million or more population), the transit mode shares for these two income groups are 11.8% and 4.5%, respectively – both higher than the national average. In contrast, for other metropolitan size categories, the transit mode shares for the lowest income group ranges from 3.7-6.2%, and that for the highest income group range from 0.5-0.8%, all lower than the above-mentioned national average of their respective income group. This implies that the transit trips are concentrated in large metropolitan areas — especially their central city neighborhoods, where most of the transit infrastructure is concentrated. Based on 2020 Census data, only 18 metropolitan areas have a population of three million or more, accounting for 36% of America's total population (U.S. Census Bureau, 2021).

Second, the lowest income group ($25,000 per year or less) was the only income group that consistently had a higher share of metropolitan-size-category-specific transit trip shares than national averages across all metropolitan size categories. For instance, for the metropolitan areas with 1-2.99 million population, the average transit trip share was 1.9%; the transit trip share for the lowest income group in this metropolitan size category is 6.1%; in contrast, the transit trip shares for all other income groups for this metropolitan size category range from 0.8-1.7%, all higher than 1.9%. This reinforces the previously-mentioned finding that the lowest income group was heavily reliant on transit. Third, the differences in transit shares between high-income and low-income groups were much smaller in large metropolitan areas than in small ones. For instance, for the largest metropolitan area group (three million or more), the transit trip share of the lowest income group was 11.8%, while that of the highest income group was 4.5% — a ratio of 2.6. In contrast, for the smallest metropolitan area group (less than 250,000), the transit trip share of the lowest income group was 5.2%, while that of the highest income group was 0.52% — a ratio of 9.9. In other words, America's large cities have a more balanced socioeconomic mix of transit users, although we caution that Table 8 shows bus riders were more likely to have a

lower socioeconomic status and rail riders were more likely to have a higher socioeconomic status. Fourth, the urban transit mode share increased from 2.8% in 2009 to 3.3% in 2017, although the differences in household income and metropolitan statistical area (MSA) categories prevent us from conducting more detailed comparisons.

**Table 10 Transit's Mode Share by Urban Size and Household Income**

| Metropolitan Area Population | Household Income | | | | | |
| --- | --- | --- | --- | --- | --- | --- |
| | Less than $25,000 | $25,000 to $49,999 | $50,000 to $74,999 | $75,000 to $99,999 | $100,000 and over | All |
| Less than 250,000 | 5.16 | 0.68 | 1.10 | 0.59 | 0.52 | 1.79 |
| 250,000–499,999 | 3.74 | 1.36 | 0.55 | 1.61 | 0.58 | 1.56 |
| 500,000–999,999 | 5.94 | 1.03 | 0.90 | 0.39 | 0.74 | 1.80 |
| 1,000,000–2,999,999 | 6.15 | 1.69 | 0.94 | 0.69 | 0.81 | 1.94 |
| 3 million or more | 11.78 | 5.61 | 4.91 | 4.98 | 4.48 | 5.75 |
| Not in MSA or CMSA | 1.41 | 0.19 | 0.04 | 1.55 | 0.26 | 0.69 |
| Nation | 6.79 | 2.61 | 2.40 | 2.43 | 2.63 | 3.31 |

Source: Calculated by the authors from the 2017 NHTS.
Note: In order to isolate urban travel, the sample was limited to residents of urban areas and trips of 75 miles or less.

As to the shares of other modes by income group indicated in Table 9, the lowest income group had a slightly higher share of non-motorized trips (19.7%) than higher income groups (12.2–14.1%). Such pattern was mainly attributed to the variations in walking trips: the share of walking trips for the lowest income group (18.1%) is considerably higher than that for other income groups (11.2-12.9%), while the shares of bicycle trips across different income groups are more similar (0.9-1.5%). Finally, we see an increase in taxicab trips from 0.2% in the 2009 NHTS to 0.6% in the 2017 NHTS. Also, the shares of taxicab trips across different income groups are at higher parity for the 2017 NHTS than the 2009 NHTS: in the 2017 survey, the income-group specific taxicab trip shares range from 0.3% to 0.7%, while in the 2009 survey, the shares range from less than 0.1% to 0.7%[1]. One explanation is that trips using ride-hailing apps (e.g., Uber or Lyft) are counted as taxicab trips in the 2017 NHTS, but these were not available options when the 2009 NHTS was conducted. For instance, for the year 2017 alone, Uber provided 4 billion rides globally, and Lyft provided 37 million rides to an almost-exclusive U.S. market – both companies were close-to non-exist in 2009 (Bhiuyan, 2018; Carson, 2018). Hence, it is possible that the increase in the total shares of taxicab trips as well as the higher parity across different income groups are attributable to the higher availability, affordability, and convenience of the ride-hailing industry (Brown, 2019; Leistner and Steiner, 2017).

Table 11 also examines the relationship between income and mode shares, in a similar but slightly different way from Table 9. The latter tabulates the mode share of each income group, whereas Table 11 examines the share of trip frequencies across different income groups for a specific mode. About one-third of automobile trips — solo driving and carpooling — were made by the highest income group. For transit trips, the lowest income group (less than $25,000) and the highest income group ($100,000 and more) make the most transit trips, with trip shares of 36.7% and 26.2%, respectively. However, a closer look reveals that the lowest income group made more than half of the total bus trips (50.9%), and the highest income group utilized subway/light rail/streetcar and commuter rail trips (41.0% and 64.6%, respectively) the most. Similarly, the highest and lowest income groups also made the most taxicab trips, with shares of 40.2% and 20.0%, respectively. Many in the lowest income group are from zero-vehicle households and need to use taxicab services meet their mobility needs, and the introduction of ride-hailing services provides a more affordable way to do so. The highest income group is less sensitive to prices but also likes to use taxicabs or ride-hailing services for convenience.

---

[1] We acknowledge that the income group cutoffs of the 2009 NHTS is slightly different from the 2017 NHTS.

**Table 11 Income Distribution of Each Mode's Users (percentage composition by income class)**

| Mode of Transportation | Household Income | | | | | |
|---|---|---|---|---|---|---|
| | Less than $25,000 | $25,000 to $49,999 | $50,000 to $74,999 | $75,000 to $99,999 | $100,000 and over | All |
| Total Auto | 15.6 | 20.7 | 16.5 | 13.7 | 33.5 | 100 |
|     SOV[a] | 14.1 | 21.5 | 17.4 | 13.9 | 33.1 | 100 |
|     HOV[b] | 16.9 | 20.0 | 15.7 | 13.5 | 33.8 | 100 |
| Total Transit | 36.7 | 15.8 | 11.6 | 9.7 | 26.2 | 100 |
|     Bus Transit[c] | 50.9 | 17.3 | 9.8 | 7.8 | 14.1 | 100 |
|     Subway/Light Rail/Streetcar[d] | 15.1 | 14.5 | 15.8 | 13.7 | 41.0 | 100 |
|     Commuter Rail[e] | 12.7 | 8.3 | 7.4 | 7.0 | 64.6 | 100 |
| Total Non-motorized | 24.7 | 17.5 | 13.9 | 11.3 | 32.6 | 100 |
|     Walk | 24.7 | 17.7 | 13.9 | 11.2 | 32.5 | 100 |
|     Bicycle | 23.9 | 15.5 | 14.8 | 12.4 | 33.4 | 100 |
| School Bus | 27.2 | 17.5 | 14.0 | 12.4 | 28.8 | 100 |
| Taxicab | 20.0 | 11.0 | 18.1 | 10.6 | 40.2 | 100 |
| Other | 26.7 | 21.9 | 19.8 | 6.1 | 25.4 | 100 |
| All | 17.9 | 20.0 | 16.0 | 13.2 | 33.0 | 100 |
| Overall Sample Distribution | | | | | | |
|     Households | 23.6 | 22.2 | 16.1 | 12.2 | 26.1 | 100 |
|     Persons | 19.9 | 20.3 | 16.0 | 12.9 | 31.0 | 100 |
|     Trips | 17.5 | 19.6 | 15.6 | 12.9 | 32.3 | 100 |

Source: Calculated by the authors from the 2017 NHTS.
Notes: In order to isolate urban travel, the sample was limited to residents of urban areas and trips of 75 miles or less.
a. SOV (single-occupancy vehicle) includes vehicles with driver and no passengers.
b. HOV (high-occupancy vehicle) includes vehicles with two or more occupants.
c. Bus transit includes public bus, commuter bus, private/shuttle bus- and city-to-city bus.
d. Subway/light rail/streetcar also includes elevated rail.
e. Commuter rail includes suburban/regional rail systems and short-distance service provided by Amtrak.

### 3.3.3 Income and time-of-day of travel

Table 12 examines trip frequency shares by income and time of day. As the table indicates, for higher income groups, the share of peak-hour trips was higher than that of non-peak-hour trips; for lower income groups, the share of non-peak-hour trips was higher than that of peak-hour trips. As to specific modes, the highest income group made 31.4% of the total peak-hour transit trips but only 22.1% of the non-peak-hour transit trips, while the lowest income group made only 28.7% of the peak-hour transit trips but 43.0% of the total non-peak-hour transit trips. For taxicab trips, the peak-hour and non-peak-hour trip shares were comparable for the lowest income group; but for the highest income group, the peak-hour trip share (41.9%) was higher than the non-peak-hour one (35.1%). These facts imply that many high-income people take transit for commuting purposes and taxicabs for social and shopping trips, while many low-income people are just the opposite, taking transit for social and shopping purposes and taking taxicabs for commuting purposes.

**Table 12 Peak vs. Off-Peak Travel by Income Class (percentage distribution of each mode's users by time of day and income)[a]**

| Mode of Transportation | Household Income | | | | | |
| --- | --- | --- | --- | --- | --- | --- |
| | Less than $25,000 | $25,000 to $49,999 | $50,000 to $74,999 | $75,000 to $99,999 | $100,000 and over | All |
| Total Auto | | | | | | |
|   Peak | 14.1 | 20.2 | 16.2 | 14.1 | 35.4 | 100 |
|   Off-Peak | 16.3 | 21.0 | 16.6 | 13.5 | 32.6 | 100 |
| Total Transit | | | | | | |
|   Peak | 28.7 | 15.1 | 12.8 | 12.0 | 31.4 | 100 |
|   Off-Peak | 43.0 | 16.3 | 10.7 | 7.9 | 22.1 | 100 |
| Bus Transit[b] | | | | | | |
|   Peak | 43.3 | 17.1 | 11.9 | 9.4 | 18.2 | 100 |
|   Off-Peak | 56.1 | 17.4 | 8.4 | 6.8 | 11.3 | 100 |
| Subway/Light Rail/Streetcar[c] | | | | | | |
|   Peak | 11.8 | 14.2 | 15.1 | 17.5 | 41.3 | 100 |
|   Off-Peak | 18.2 | 14.8 | 16.4 | 10.0 | 40.7 | 100 |
| Commuter Rail[d] | | | | | | |
|   Peak | 6.5 | 6.0 | 8.2 | 5.5 | 73.7 | 100 |
|   Off-Peak | 21.1 | 11.3 | 6.3 | 9.1 | 52.2 | 100 |
| Taxicab, Including Uber/Lyft | | | | | | |
|   Peak | 20.9 | 18.7 | 16.4 | 9.0 | 35.1 | 100 |
|   Off-Peak | 19.7 | 8.6 | 18.7 | 11.1 | 41.9 | 100 |
| All Modes | | | | | | |
|   Peak | 16.7 | 19.6 | 15.7 | 13.5 | 34.5 | 100 |

|   | Off-Peak | 18.4 | 20.2 | 16.1 | 13.0 | 32.3 | 100 |
|---|---|---|---|---|---|---|---|
| All Modes and All Incomes | | | | | | | |
| | Peak | | | | | | 32.3 |
| | Off-Peak | | | | | | 67.7 |

Source: Calculated by the authors from the 2017 NHTS.
Notes: In order to isolate urban travel, the sample was limited to residents of urban areas and trips of 75 miles or less.
a. Peak period was defined as 6 to 9 a.m. and 4 to 7 p.m. on weekdays; off-peak included all other times.
b. Bus transit includes public bus, commuter bus, private/shuttle bus, and city-to-city bus.
c. Subway/light rail/streetcar also includes elevated rail.
d. Commuter rail includes suburban/regional rail systems and short-distance service provided by Amtrak.

### 3.3.4 Housing tenure and travel modes

The final economic factor that we examined is housing tenure, as shown in Table 13. Compared with the 2009 NHTS (Renne and Bennett, 2014), the shares of automobile trips for both owners and renters decreased in the 2017 NHTS. Specifically, the shares of automobile trips for owners and renters in 2009 were 86.1% and 72.2%, respectively, dropping to 84.7% and 71.0%, respectively, in 2017. However, a closer look reveals that the drops for owner and renters have different reasons. For owners, the share of solo driving trips decreased (from 42.3% to 40.1%), and the share of carpooling trips has increased (from 43.8% to 44.7%); renters exhibit the opposite trend: the share of solo driving trips increased (from 30.5% to 32.1%), and the share of carpooling trips decreased (from 41.6% to 38.8%). One explanation is that there are more highly educated, high-income people becoming renters, and they drive for social and shopping trips but prefer using transit rather than carpooling for commute trips. With respect to transit trips, the shares for both owners and renters have increased (from 1.1% to 1.7% for owners and from 5.8% to 6.1% for renters). Also, the share of taxicabs increased for owners from 0.1% in 2009 to 0.3% in 2017, and for renters from 0.5% in 2009 to 1.1% to 2017. These facts indicate that transportation network companies, such as Uber and Lyft have become particularly popular among urban residents, especially for renters.

**Table 13 Variations in Modal Choice by Housing Tenure**

| Mode of Transportation | Housing Tenure | | |
|---|---|---|---|
| | Own | Rent | Total |
| Percent of All Households | 60.1 | 39.9 | 100 |
| Percent of All Trips | 63.6 | 36.4 | 100 |
| | | | |
| Total Auto | 84.7 | 71.0 | 79.7 |
|     SOV[a] | 40.1 | 32.1 | 37.1 |
|     HOV[b] | 44.7 | 38.8 | 42.6 |
| Total Transit | 1.7 | 6.1 | 3.3 |
|     Bus Transit[c] | 0.8 | 4.1 | 2.0 |
|     Subway/Light Rail/Streetcar[d] | 0.6 | 1.8 | 1.1 |
|     Commuter Rail[e] | 0.2 | 0.2 | 0.2 |
| Total Non-motorized | 11.3 | 19.3 | 14.2 |
|     Walk | 10.4 | 17.7 | 13.1 |
|     Bicycle | 0.9 | 1.5 | 1.1 |
| School Bus | 1.6 | 2.0 | 1.7 |
| Taxicab | 0.3 | 1.1 | 0.6 |
| Other | 0.4 | 0.5 | 0.4 |
| All | 100.0 | 100.0 | 100.0 |

Source: Calculated by the authors from the 2017 NHTS.

Notes: In order to isolate urban travel, the sample was limited to residents of urban areas and trips of 75 miles or less.

a. SOV (single-occupancy vehicle) includes vehicles with driver and no passengers.
b. HOV (high-occupancy vehicle) includes vehicles with two or more occupants.
c. Bus transit includes public bus, commuter bus, private/shuttle bus, and city-to-city bus.
d. Subway/light rail/streetcar also includes elevated rail.
e. Commuter rail includes suburban/regional rail systems and short-distance service provided by Amtrak.

*3.4   Demographic factors and urban travel*

Besides varying by economic conditions, urban travel patterns also vary by demographic factors, such as ethnicity, gender, age, and lifecycle factors. This section discusses the relationships between these demographic factors and various travel patterns.

As indicated in Table 14, non-Hispanic whites are the most auto-oriented ethnicity group, with 81.5% of their trips by automobile, followed by Hispanics with 80.2% of their trips by automobile. Within automobile trips, Hispanics are higher in carpooling, as 48.0% of their total trips were carpooling trips, while the number for non-Hispanic whites is only 41.3%. As the "less auto-oriented groups," including blacks and Asians also had the highest share of transit trips. Taking a closer look reveals that blacks were the highest in bus transit trips (5.8%), and Asians were the highest in rail transit trips (2.2%). Asians also had the highest share of non-motorized trips (17.7%), and Hispanics had the lowest share of non-motorized trips (13.1%). Although the share of non-motorized trips for the whites and blacks is comparable (around 14%), whites had a much higher share of cycling trips than blacks (1.3% vs. 0.5%, respectively). Finally, although the share of taxicab trips was comparable across different racial profiles, we do see that the share of taxicab trips was lowest among whites, at 0.5%, while other racial groups are all 0.7–0.8%. Such variations across racial profiles likely reflect differences in their economic conditions, lifestyles, and immigration status (Taylor and Morris, 2015; Shin, 2017; Wang, 2015).

**Table 14 Variations in Modal Choice by Race/Ethnicity (percentage of trips by mode of transportation)**

| Mode of Transportation | Race/Ethnicity | | | |
| --- | --- | --- | --- | --- |
| | Non-Hispanic Black | Non-Hispanic Asian | Non-Hispanic White | Hispanic |
| Total Auto | 73.9 | 73.5 | 81.5 | 80.2 |
|     SOV[a] | 35.5 | 29.7 | 40.2 | 32.2 |
|     HOV[b] | 38.4 | 43.8 | 41.3 | 48.0 |
| Total Transit | 7.6 | 5.7 | 2.1 | 3.6 |
|     Bus Transit[c] | 5.8 | 3.0 | 1.0 | 2.5 |
|     Subway/Light Rail/Streetcar[d] | 1.6 | 2.2 | 0.9 | 0.9 |
|     Commuter Rail[e] | 0.2 | 0.4 | 0.2 | 0.2 |
| Total Non-motorized | 14.1 | 17.7 | 14.3 | 13.1 |
|     Walk | 13.6 | 16.4 | 13.0 | 12.3 |
|     Bicycle | 0.5 | 1.3 | 1.3 | 0.8 |
| School Bus | 3.0 | 2.0 | 1.3 | 2.0 |
| Taxicab | 0.7 | 0.8 | 0.5 | 0.7 |
| Other | 0.7 | 0.3 | 0.4 | 0.4 |
| All | 100 | 100 | 100 | 100 |
| Overall Sample Distribution[f] | | | | |
|     Percent of All Individuals | 13.5 | 6.2 | 56.4 | 19.5 |
|     Percent of Total Trips | 12.5 | 5.5 | 59.4 | 18.4 |

Source: Calculated by the authors from the 2017 NHTS.
Notes: In order to isolate urban travel, the sample was limited to residents of urban areas and trips of 75 miles or less.
a. SOV (single-occupancy vehicle) includes vehicles with driver and no passengers.
b. HOV (high-occupancy vehicle) includes vehicles with two or more occupants.
c. Bus transit includes public bus, commuter bus, private/shuttle bus, and city-to-city bus.
d. Subway/light rail/streetcar also includes elevated rail.
e. Commuter rail includes suburban/regional rail systems and short-distance service provided by Amtrak.
f. Rows do not total 100% because some racial and ethnic categories are not shown.

As to gender differences, females had slightly higher shares of automobile trips than males (80.6% vs. 78.7%); specifically, we see females had a higher share of carpools than males (45.2% vs. 39.6%), and males had a higher share of solo driving than females (39.1% vs. 35.4%). One explanation is that females are more likely to be the ones dropping off and picking up children at school (McDonald, 2008). Males have higher shares of bicycle trips than females (1.6% vs. 0.7%), which corresponds to an international study showing that females are less likely to ride bicycles in cities with cycling rates lower than 7% (Goel et al., 2022). Except for car and non-motorized trips, shares in other modes are comparable between males and females. Compared with the 2009 NHTS, we noticed drops in the automobile mode shares for both males and females and increases in transit mode shares for both genders. There were also increases in taxicab trips for both genders: the taxicab mode shares for males and females in 2009 were 0.2% and 0.3%, respectively, while in 2017, the shares increased to 0.6% and 0.5%, respectively. That said, it is worth noting that the share of taxicabs in 2009 (before Uber/Lyft) was slightly higher for females, but the share of taxicabs in 2017 (with Uber/Lyft) was slightly lower for females. One possible explanation is that safety concerns about ride-hailing services are relatively greater for females (IFC, 2018).

**Table 15 Variations in Modal Choice by Gender (percentage of trips by means of transportation)**

| Mode of Transportation | Gender | | |
|---|---|---|---|
| | **Male** | **Female** | **All** |
| Total Auto | 78.7 | 80.6 | 79.7 |
|     SOV[a] | 39.1 | 35.4 | 37.1 |
|     HOV[b] | 39.6 | 45.2 | 42.6 |
| Total Transit | 3.4 | 3.3 | 3.3 |
|     Bus Transit[c] | 2.0 | 2.0 | 2.0 |
|     Subway/Light Rail/Streetcar[d] | 1.1 | 1.0 | 1.1 |
|     Commuter Rail[e] | 0.3 | 0.2 | 0.2 |
| Total Non-motorized | 14.8 | 13.7 | 14.2 |
|     Walk | 13.1 | 13.1 | 13.1 |
|     Bicycle | 1.6 | 0.7 | 1.1 |
| School Bus | 2.0 | 1.5 | 1.7 |
| Taxicab | 0.6 | 0.5 | 0.6 |
| Other | 0.6 | 0.3 | 0.4 |
| All | 100.0 | 100.0 | 100.0 |

Source: Calculated by the authors from the 2017 NHTS.
Notes: In order to isolate urban travel, the sample was limited to residents of urban areas and trips of 75 miles or less.
a. SOV (single-occupancy vehicle) includes vehicles with driver and no passengers.
b. HOV (high-occupancy vehicle) includes vehicles with two or more occupants.
c. Bus transit includes public bus, commuter bus, private/shuttle bus, and city-to-city bus.
d. Subway/light rail/streetcar also includes elevated rail.
e. Commuter rail includes suburban/regional rail systems and short-distance service provided by Amtrak.

Table 16 shows that mobility levels — measured by trip frequency and distance — were relatively lower for the young and the old. For trip frequencies, the biggest increase occurs at the age of 25 (from 3.1 trips per person, per day for ages 16–24, to 3.8 for ages 25–39), whereas the average daily trip frequency is quite stable for the age groups from 25 to 74 (3.8-3.9 trips per person, per day); the value drops to 3.2 trips per day for the 75-and-older age group. For trip distances, there is a clear Inverse-U relationship with age: the peak is in the 25–39 age group (with 24.3 miles) and 40–64 age groups (24.7 miles), followed by early adulthood, youth, and elderly (18.4 miles for age 16–24, and 18.8 miles for age 65–74). The oldest and youngest groups were the lowest (13.9 miles for age 75 and older, 12.7 miles for age 5–15). With respect to mode shares, as indicated by Table 17, although older adults have lower trip distance and frequencies than younger age groups, the former's mode share is more auto-oriented. For instance, for the 65–74 and 75 and older age groups, more than 80% of their trips were made by automobiles, whereas the auto mode shares for the 16–24 and 25–39 age groups were under 80%. Also, the share of solo driving is larger than that of carpooling for older adults, while the share of carpooling is higher than that of solo driving for many of the younger age groups. In addition, the share of transit trips for older adults is lower than that of the younger age groups, showing that when older adults reduce their automobile trip frequency and distances, they do not use transit to compensate for their reduced mobility. One explanation is that older adults prefer the flexibility and freedom of cars and are deterred by transit's inconvenience and lack of senior-friendly facilities (Loukaitou-Sideris, Wachs and Pinski, 2019; Wang, 2022). Further, the share of taxicab trips for older adults was the lowest across all age groups, which is likely to be a combination of lower smartphone usage and safety concerns about ride-hailing services (Leistner and Steiner, 2017).

**Table 16 Impact of Age on Mobility Levels**

| Age Group | Trips per Day, per Person | Miles Travelled per Day, per Person | |
|---|---|---|---|
| | | Unadjusted | Adjusted |
| 5 to 15 | 3.0 | 12.7 | 14.0 |
| 16 to 24 | 3.1 | 18.4 | 20.1 |
| 25 to 39 | 3.8 | 24.3 | 26.5 |
| 40 to 64 | 3.9 | 24.7 | 27.0 |
| 65 to 74 | 3.8 | 18.8 | 20.8 |
| 75 & over | 3.2 | 13.9 | 15.4 |
| All | 3.6 | 20.8 | 22.8 |

Source: Calculated by the authors from the 2017 NHTS.
Note: In order to isolate urban travel, the sample was limited to residents of urban areas and trips of 75 miles or less.

**Table 17 Variations in Mode Share by Age (percentage of trips by means of transportation)**

| Mode of Transportation | | Age | | | | | |
|---|---|---|---|---|---|---|---|
| | | 5 to 15 | 16 to 24 | 25 to 39 | 40 to 64 | 65 to 74 | 75 & over |
| Total Auto | | 70.7 | 77.0 | 79.1 | 82.8 | 81.7 | 83.2 |
| | SOV | 0.7 | 36.5 | 39.4 | 45.3 | 43.8 | 41.9 |
| | HOV | 70.0 | 40.6 | 39.8 | 37.5 | 37.9 | 41.4 |
| Total Transit | | 1.4 | 4.7 | 4.0 | 3.3 | 3.0 | 2.4 |
| | Public or Commuter Bus | 1.0 | 3.1 | 2.0 | 2.0 | 2.2 | 2.1 |
| | Subway/Light Rail/Streetcar | 0.3 | 1.3 | 1.7 | 1.0 | 0.7 | 0.3 |
| | Commuter Rail | 0.0 | 0.3 | 0.3 | 0.3 | 0.1 | 0.0 |
| Total Non-motorized | | 15.1 | 15.2 | 15.6 | 12.8 | 14.3 | 13.6 |
| | Walk | 13.3 | 13.9 | 14.2 | 11.9 | 13.6 | 13.3 |
| | Bicycle | 1.9 | 1.3 | 1.4 | 0.9 | 0.7 | 0.4 |
| School Bus | | 12.3 | 1.8 | 0.1 | 0.1 | 0.1 | 0.0 |
| Taxicab | | 0.2 | 0.7 | 0.9 | 0.5 | 0.4 | 0.2 |
| Other | | 0.3 | 0.5 | 0.3 | 0.5 | 0.4 | 0.5 |
| All | | 100.0 | 100.0 | 100.0 | 100.0 | 100.0 | 100.0 |

Source: Calculated by the authors from the 2017 NHTS.
Notes: In order to isolate urban travel, the sample was limited to residents of urban areas and trips of 75 miles or less.
a. SOV (single-occupancy vehicle) includes vehicles with driver and no passengers.
b. HOV (high-occupancy vehicle) includes vehicles with two or more occupants.
c. Bus transit includes public bus, commuter bus, private/shuttle bus, and city-to-city bus.
d. Subway/light rail/streetcar also includes elevated rail.
e. Commuter rail includes suburban/regional rail systems and short-distance service provided by Amtrak.

The final demographic factor that we examine is life-cycle classification, a factor closely related to — but not identical with — age (Garikapati et al., 2016; Wang and Wang, 2021). Although the average trip frequency and distance do not differ much across life-cycle types (Table 18), the mode shares do (Table 19). Specifically, the biggest difference comes from the "no children" group vs. the groups having young children: the "no children" group has an automobile share of 75.9%, while the auto trip shares among the "having young children" groups are at least 80%. The share of solo driving trips was 46.2% for the "no children" group, while the shares of the "youngest child 0–5" and "youngest child 6–15" groups were both below 30%, indicating families engage in many carpooling trips when their children were not eligible for a driver's license (most states issue driver's licenses at 16 years old). As to transit trips, the trip share for the "no children" group was 5.1%, and all other "having young children" groups' trip shares were about 2–3%. In sum having children — especially children younger than 16 years old — was the biggest difference maker for travel behaviors (Garikapati et al., 2016; McDonald, 2008).

**Table 18 Variations in Mobility Levels by Life-Cycle Classifications**

| Life-Cycle Classifications of Household | Trips per Day, per Person | Miles Travelled per Day, per Person | |
| --- | --- | --- | --- |
| | | Unadjusted | Adjusted |
| No children | 3.8 | 23.2 | 25.3 |
| Youngest child 0–5 | 3.5 | 20.0 | 21.9 |
| Youngest child 6–15 | 3.6 | 20.1 | 22.1 |
| Youngest child 16–21 | 3.3 | 21.7 | 23.7 |
| Retired, no children[a] | 3.5 | 18.2 | 20.1 |
| All | 3.6 | 20.8 | 22.8 |

Source: Calculated by the authors from the 2017 NHTS.
Notes: In order to isolate urban travel, the sample was limited to residents of urban areas and trips of 75 miles or less.
a. "Retired, no children" includes having children (21 or younger) who are not living in the same household and having children older than 21 who are living in the same household.

**Table 19 Variations in Modal Choice by Life-Cycle Classifications**

| Mode of Transportation | Life-Cycle Classifications of Household | | | | |
|---|---|---|---|---|---|
| | No children | Youngest child 0–5 | Youngest child 6–15 | Youngest child 16–21 | Retired, no children[a] |
| Total Auto | 75.9 | 82.0 | 80.0 | 83.1 | 82.3 |
|     SOV | 46.2 | 28.0 | 25.5 | 44.5 | 42.0 |
|     HOV | 29.8 | 53.9 | 54.5 | 38.6 | 40.3 |
| Total Transit | 5.1 | 2.2 | 2.1 | 3.2 | 2.8 |
|     Bus Transit | 2.9 | 1.4 | 1.3 | 1.7 | 2.2 |
|     Subway/Light Rail/Streetcar | 1.9 | 0.6 | 0.6 | 1.2 | 0.5 |
|     Commuter Rail | 0.3 | 0.2 | 0.2 | 0.3 | 0.1 |
| Total Non-motorized | 17.2 | 12.5 | 12.5 | 11.7 | 14.0 |
|     Walk | 15.8 | 11.5 | 11.4 | 10.7 | 13.3 |
|     Bicycle | 1.5 | 1.0 | 1.1 | 1.0 | 0.7 |
| School Bus | 0.1 | 2.7 | 4.7 | 1.1 | 0.0 |
| Taxicab | 1.1 | 0.3 | 0.3 | 0.3 | 0.3 |
| Other | 0.5 | 0.3 | 0.4 | 0.5 | 0.5 |
| All | 100 | 100 | 100 | 100 | 100 |

Source: Calculated by the authors from the 2017 NHTS.
Notes: In order to isolate urban travel, the sample was limited to residents of urban areas and trips of 75 miles or less.
a. "No children" includes having children (21 or younger) who are not living in the same household and having children older than 21 who are living in the same household.

## 4 Conclusions and policy implications

The most noticeable takeaway of the 2017 NHTS is that although private automobiles continue to be the dominant travel mode for Americans, we saw a slight decline in the share of car trips after its peak in 2001, especially in urban areas. For the whole U.S., the share of automobile travel dropped from 86.4% in 2001 to 83.6% in 2009 and further to 82.6% in 2017 (Table 2). We saw a similar trend in automobile share in commute trips: which was 87.9% in the 2000 census, 86.1% in the 2009 ACS one-year estimate, and 85.3% in the 2017 ACS one-year estimate (Table 1). For urban America, the share of auto travel peaked at 85.9% in 2001, dropped to 81.9% in 2009, and further dropped to 79.7% in 2017 (Pucher and Renne, 2003; Renne and Bennett, 2014). A detailed look shows that the share of solo driving trips increased from 2001 to 2009 (37.3% to 38.6%) but decreased from 2009 to 2017 (38.6% to 37.1%), while the share of carpooling has consistently decreased from 48.6% in 2001 to 43.3% in 2009 and 42.6% in 2017. Although we can attribute the decrease from 2001 to 2009 to the economic recession, the continuation of this trend, based on 2017 data, may give us some cautious optimism that America's urban travel may experience a similar "peak and drop" trend in automobile travel as other advanced economies (Millard-Ball and Schipper, 2011; Wang and Akar, 2020). However, the dawn of ride-hailing between 2009 and 2017 could have had an impact on carpooling, in favor of trips using services like Uber and Lyft (Brown, 2019; Dong, 2020).

    In addition, we see steady increases in the share of transit trips for urban travel. The share of total transit trips was 1.7% in 2001, increased to 2.8% in 2009, and further increased to 3.3% in 2017. Specifically, since the share of commuter rail trips remained relatively stable over time (0.1% in 2001, 0.2% in both 2009 and 2017), most of the increases come from the increasing share of bus transit trips as well as light rail/subway/streetcar trips (1.6% in 2001, 2.6% in 2009, and 3.1% in 2017). Perhaps the dawn of micromobility services between 2009 and 2017 made transit more appealing since last-mile connections became easier (Beak et al., 2021; Hong, Jiang and Lee, 2023). Although we are unable to further explore the temporal changes in specific shares (e.g., bus vs. rail) due to differences in the mode and income classifications between the 2017 NHTS and its predecessors, we see the increases in transit trip shares are consistent across housing ownership, ethnicity, and gender. Similarly, the share of non-motorized trips increased from 10.4% in 2001 to 12.9% in 2009 and 14.2% in 2017, with most of this increase coming from walking trips (9.2% in 2001, 11.8% in 2009, and 13.1% in 2017). In addition, we also see an increase in the share of taxicabs, which rose from 0.1% in 2001 to 0.2% in 2009 and 0.6% in 2017. Since the 2017 survey included ride-hailing trips (e.g., Uber/Lyft) as a part of "taxi," we

cannot precisely determine how much of the increase can be attributed to the rise of the ride-hailing industry. Nonetheless, ride-hailing services provide a relatively affordable and flexible way to get around without taking private automobiles — especially for carless households in low-income neighborhoods (Brown, 2019; Dong, 2020).

Many different factors may be contributing to this general trend of "peak and decline" for the share of automobile trips and the steady increases in transit, non-motorized, and taxicab trips in urban America. First, the continuous support from federal, state, and local sources to build transit infrastructure and encourage transit-oriented development has enabled urban residents to reduce car use and increase transit use (Spears, Boarnet and Houston, 2017; Lowe, 2013; Renne and Appleyard, 2019). Second, the past two decades have seen the rise and growth of the IT industry as well as other creative industries in urban neighborhoods and college towns with robust transit and non-motorized infrastructures; such "urban clustering" of industrial growth not only causes growth in the populations of transit- and non-motorized-friendly communities but also brings about lifestyle changes in terms of living, traveling, and shopping (Alder and Florida, 2021; Blumenberg et al., 2019; Lee, 2020; Wang, 2019). Third, higher gasoline price reduces the frequency and distance of driving (Bastian, Borjesson and Eliasson, 2016; Stapleton, Sorrell and Schwanen, 2017); and the fuel price is likely to remain high in the near future due to political and policy reasons. Fourth, the rise of online retailers such as Amazon and Uber Eats has reduced the demand for shopping trips and evening dining trips due to convenience and often lower prices (Cao, 2012; Le, Carrel and Shah, 2022). Although we observe a decline in passenger travel by automobiles, we do see a steady increase in freight truck ton-miles over the last decade (Bureau of Transportation Statistics, 2020). Fifth, the rise of shared mobility — both ride-hailing, bike and scooter sharing — might also have helped attract urban residents to non-automobile modes, since last-mile accessibility is easier. As discussed in the previous paragraph, we have seen the increase in taxicab trips for owners and renters, males and females, and across different ethnicities. It is very likely that ride-hailing services are the main mechanism of this universal rise, as they provide a more convenient and reasonably affordable way to travel (Brown, 2019; Dong, 2020). On the other hand, the prosperity of bike- and scooter-share programs across U.S. cities can help solve the first-mile/last-mile problem for potential transit takers, making transit more attractive than driving (Fukushige, Fitch and Handy, 2021; Martin and Xu, 2022). Sixth, working from home and flexible working schedules have been gaining popularity in American cities. As shown in Table 1, the share of Americans working from home increased from 3.3% in 2000 to 4.3% in 2009, and 5.2% in 2017, and post-covid American cities will see a further increases in the share of working-from-home arrangements (Li and Wang, 2022). Although it is still unclear whether people will make more trips to compensate for their reduced commute trips, it is worth noting that commute trips are more likely to be solo driving and long-distance trips, while non-commute trips are more likely to be transit and non-motorized trips of shorter distances (Boarnet, 2011; McDonald, 2015).

Most, if not all, of the six above-mentioned factors should continue into the near future: the transit system continues to expand, central city areas will still be IT-incubators with residents leading urban-oriented lifestyles, the fuel price may remain high, the online retail industry will keep prospering, shared mobility services will continue to be popular, and a few working-from-home days per week very likely will become the new normal in a post-pandemic world. Thus, although it is still too early to tell whether such a "peak-and-decline" trend in automobile travel will continue in urban America, the current trend provides an opportunity to encourage multimodalism in American cities. Further development of transit networks, more support of transit-oriented development (from both public and private sectors), and ongoing expansion of non-motorized transport infrastructure could help American cities increase growth in becoming "multimodal" in the future rather than reverting to a "car-dominant" past. Such a "multimodal" future will not only increase the environmental sustainability of American cities but also boost the economic and social sustainability of providing low-income urban residents with more convenient, flexible, and affordable ways to reach educational and economic opportunities (Chetty et al., 2022; Fan, Guthrie and Levinson, 2012).

      Indeed, there are still many challenges for American cities to move toward multimodality and more equity in mobility. As this paper suggests, a general trend may mask the complex relationships between socioeconomic factors and travel pattens, and certain economic and demographic sub-groups face higher challengers than others. For instance, similar to earlier studies in this series using earlier NPTS/NHTS data, we see the lowest income group (household income less than $25,000), who were also more likely to be minorities and carless, were heavily reliant on transit – especially buses – to fulfill their mobility needs. Their daily trip frequencies and daily miles traveled were considerably lower than those of higher income groups, implying that their social activities and economic opportunities were also likely to be limited (Gobillon, Selod & Zenou, 2007; Klein, 2020). Such mobility challenges for the lowest income group were more severe in small- and mid-size cities, where the supply of transit is more limited than large cities. Moreover, the American population is rapidly aging, and more than 20% of the total population will be 65 and older by 2030. As previously discussed, most older adults heavily relied on driving and are not regular transit patrons, ride-hailing users, or cyclists. As they gradually reduce and even cease driving, providing senior-friendly, safe, and convenient alternative mobility options will be essential for maintaining their health and well-being

(Schouten et al., 2022; Loukaitou-Sideris, Wachs and Pinski, 2019). Also, the taxicab trips in the 2017 NHTS were more evenly distributed across socioeconomic groups compared with earlier surveys. Although such trends show the potential for ride-hailing to provide mobility options for those cannot afford cars or cannot drive cars, affordability issues for lower income people and acceptability issues for older adults may become barriers for equal access to TNCs (Brown et al., 2022; Leistner & Steiner, 2017). In sum, policymakers should consider the differences in travel patterns between higher-income vs. lower-income households, whites vs. non-whites, males vs. females, and young vs. old to ensure safe and convenient mobility options are provided equitably rather than catering to certain groups at the cost of others.

**Acknowledgement**

This research is supported by the Singapore Ministry of Education Academic Research Fund Tier 1. We would like to thank the helpful comments and suggestions by the Editor and Referees. All errors and omissions are the responsibility of the authors.

# Appendix

**Table A1 State-level variations in Modal Shares for Transit, Walking and Bicycling (percentage of trips by mode)**

| State | Transit | | | | Non-motorized | | |
|---|---|---|---|---|---|---|---|
| | Total | Bus Transit | Subway/Light Rail/Streetcar | Commuter rail | Total | Walk | Bicycle |
| Alabama | 1.33 | 1.33 | 0.00 | 0.00 | 5.57 | 5.47 | 0.10 |
| Alaska | 1.95 | 1.25 | 0.69 | 0.00 | 12.45 | 10.16 | 2.29 |
| Arizona | 1.75 | 1.38 | 0.37 | 0.00 | 11.87 | 10.24 | 1.63 |
| Arkansas | 0.16 | 0.16 | 0.00 | 0.00 | 8.66 | 7.99 | 0.67 |
| California | 3.19 | 2.25 | 0.71 | 0.22 | 15.45 | 14.13 | 1.32 |
| Colorado | 2.24 | 2.07 | 0.17 | 0.00 | 16.77 | 13.79 | 2.99 |
| Connecticut | 2.94 | 2.42 | 0.05 | 0.47 | 11.15 | 10.25 | 0.89 |
| Delaware | 1.58 | 1.58 | 0.00 | 0.00 | 14.21 | 13.78 | 0.43 |
| District of Columbia | 17.64 | 9.59 | 7.88 | 0.17 | 40.83 | 37.54 | 3.28 |
| Florida | 1.13 | 1.01 | 0.08 | 0.04 | 11.61 | 10.35 | 1.25 |
| Georgia | 2.08 | 1.49 | 0.47 | 0.12 | 11.40 | 10.70 | 0.70 |
| Hawaii | 2.62 | 2.62 | 0.00 | 0.00 | 16.46 | 15.34 | 1.12 |
| Idaho | 0.33 | 0.33 | 0.00 | 0.00 | 16.66 | 13.52 | 3.14 |
| Illinois | 6.43 | 2.99 | 2.41 | 1.03 | 16.77 | 16.06 | 0.71 |
| Indiana | 0.50 | 0.43 | 0.03 | 0.04 | 9.24 | 8.43 | 0.81 |
| Iowa | 1.43 | 1.43 | 0.00 | 0.00 | 11.01 | 9.71 | 1.30 |
| Kansas | 0.00 | 0.00 | 0.00 | 0.00 | 8.72 | 7.67 | 1.05 |
| Kentucky | 1.14 | 1.14 | 0.00 | 0.00 | 10.90 | 10.79 | 0.11 |
| Louisiana | 1.52 | 1.52 | 0.00 | 0.00 | 14.96 | 14.62 | 0.33 |
| Maine | 1.28 | 1.28 | 0.00 | 0.00 | 18.79 | 18.69 | 0.11 |
| Maryland | 4.17 | 2.79 | 1.30 | 0.08 | 12.26 | 11.38 | 0.88 |
| Massachusetts | 6.90 | 3.08 | 3.20 | 0.62 | 21.61 | 20.09 | 1.52 |
| Michigan | 2.18 | 2.18 | 0.00 | 0.00 | 13.40 | 11.91 | 1.48 |
| Minnesota | 3.00 | 2.45 | 0.55 | 0.00 | 12.00 | 9.59 | 2.40 |
| Mississippi | 0.96 | 0.96 | 0.00 | 0.00 | 6.11 | 5.58 | 0.53 |
| Missouri | 1.29 | 1.26 | 0.02 | 0.00 | 11.80 | 10.92 | 0.87 |
| Montana | 1.34 | 1.11 | 0.00 | 0.24 | 12.30 | 10.50 | 1.80 |
| Nebraska | 1.29 | 1.29 | 0.00 | 0.00 | 12.15 | 11.69 | 0.46 |
| Nevada | 2.84 | 2.84 | 0.00 | 0.00 | 17.53 | 16.09 | 1.44 |
| New Hampshire | 0.88 | 0.88 | 0.00 | 0.00 | 11.05 | 10.54 | 0.51 |
| New Jersey | 4.44 | 2.04 | 1.56 | 0.84 | 19.16 | 18.54 | 0.62 |
| New Mexico | 2.04 | 1.93 | 0.12 | 0.00 | 11.27 | 8.45 | 2.82 |
| New York | 13.34 | 4.40 | 7.91 | 1.03 | 28.09 | 26.62 | 1.47 |
| North Carolina | 1.48 | 1.39 | 0.09 | 0.00 | 11.44 | 10.61 | 0.84 |
| North Dakota | 0.48 | 0.48 | 0.00 | 0.00 | 7.83 | 6.73 | 1.10 |
| Ohio | 1.78 | 1.72 | 0.04 | 0.01 | 10.87 | 10.08 | 0.79 |
| Oklahoma | 0.93 | 0.87 | 0.06 | 0.00 | 8.07 | 7.91 | 0.16 |
| Oregon | 2.37 | 1.79 | 0.58 | 0.00 | 15.01 | 12.37 | 2.64 |
| Pennsylvania | 3.61 | 2.66 | 0.69 | 0.25 | 16.49 | 15.78 | 0.70 |
| Rhode Island | 2.01 | 1.88 | 0.07 | 0.05 | 17.12 | 14.43 | 2.69 |
| South Carolina | 0.84 | 0.76 | 0.08 | 0.00 | 8.84 | 8.21 | 0.64 |
| South Dakota | 0.00 | 0.00 | 0.00 | 0.00 | 8.53 | 7.61 | 0.92 |
| Tennessee | 1.43 | 1.39 | 0.03 | 0.01 | 9.35 | 8.46 | 0.89 |
| Texas | 1.62 | 1.36 | 0.22 | 0.03 | 9.15 | 8.29 | 0.86 |
| Utah | 1.03 | 0.47 | 0.43 | 0.13 | 12.05 | 11.56 | 0.49 |
| Vermont | 1.23 | 1.23 | 0.00 | 0.00 | 20.42 | 18.01 | 2.40 |
| Virginia | 3.38 | 2.04 | 1.12 | 0.22 | 13.75 | 13.32 | 0.43 |
| Washington | 4.87 | 4.20 | 0.67 | 0.00 | 14.33 | 13.02 | 1.32 |
| West Virginia | 0.97 | 0.97 | 0.00 | 0.00 | 5.84 | 5.84 | 0.00 |
| Wisconsin | 2.12 | 2.08 | 0.03 | 0.00 | 12.70 | 10.98 | 1.71 |
| Wyoming | 0.59 | 0.59 | 0.00 | 0.00 | 8.09 | 5.75 | 2.34 |

Source: Calculated by the authors from the 2017 NHTS.

Note: In order to isolate urban travel, the sample was limited to residents of urban areas and trips of 75 miles or less.

**Table A2 MSA-level variations in Modal Shares for Transit, Walking and Bicycling (percentage of trips by mode, only for MSA with 1 million or more population)**

| MSA | Transit | | | | Non-motorized | | |
|---|---|---|---|---|---|---|---|
| | Total | Bus Transit | Subway/ Light Rail/ Streetcar | Commuter rail | Total | Walk | Bicycle |
| Atlanta-Sandy Springs-Roswell, GA | 2.43 | 1.75 | 0.68 | 0.18 | 11.27 | 10.87 | 0.40 |
| Austin-Round Rock, TX | 1.71 | 1.68 | 0.04 | 0.01 | 13.34 | 11.96 | 1.38 |
| Baltimore-Columbia-Towson, MD | 4.72 | 4.63 | 0.09 | 0.03 | 12.94 | 11.98 | 0.97 |
| Birmingham-Hoover, AL | 0.60 | 0.60 | 0.00 | 0.00 | 4.04 | 4.04 | 0.00 |
| Boston-Cambridge-Newton, MA-NH | 6.78 | 2.65 | 4.13 | 0.70 | 22.41 | 20.97 | 1.44 |
| Buffalo-Cheektowaga-Niagara Falls, NY | 3.77 | 3.51 | 0.26 | 0.00 | 14.62 | 13.51 | 1.11 |
| Charlotte-Concord-Gastonia, NC-SC | 1.37 | 0.97 | 0.40 | 0.00 | 10.83 | 10.20 | 0.62 |
| Chicago-Naperville-Elgin, IL-IN-WI | 6.17 | 3.09 | 3.08 | 1.32 | 18.67 | 17.84 | 0.83 |
| Cincinnati, OH-KY-IN | 0.35 | 0.35 | 0.00 | 0.00 | 8.04 | 7.77 | 0.28 |
| Cleveland-Elyria, OH | 3.71 | 3.63 | 0.09 | 0.07 | 13.25 | 12.48 | 0.77 |
| Columbus, OH | 2.12 | 2.12 | 0.00 | 0.00 | 9.81 | 8.91 | 0.90 |
| Dallas-Fort Worth-Arlington, TX | 1.40 | 0.98 | 0.42 | 0.11 | 8.70 | 8.09 | 0.60 |
| Denver-Aurora-Lakewood, CO | 2.32 | 2.03 | 0.29 | 0.00 | 17.09 | 14.74 | 2.35 |
| Detroit-Warren-Dearborn, MI | 1.78 | 1.78 | 0.00 | 0.00 | 12.10 | 10.60 | 1.50 |
| Grand Rapids-Wyoming, MI | 0.39 | 0.39 | 0.00 | 0.00 | 19.40 | 18.06 | 1.35 |
| Hartford-West Hartford-East Hartford, CT | 1.11 | 1.11 | 0.00 | 0.00 | 12.20 | 11.50 | 0.70 |
| Houston-The Woodlands-Sugar Land, TX | 2.16 | 1.77 | 0.40 | 0.01 | 10.26 | 9.19 | 1.07 |
| Indianapolis-Carmel-Anderson, IN | 0.33 | 0.33 | 0.00 | 0.00 | 10.52 | 9.35 | 1.16 |
| Jacksonville, FL | 0.75 | 0.75 | 0.00 | 0.00 | 13.61 | 13.61 | 0.00 |
| Kansas City, MO-KS | 1.06 | 1.06 | 0.00 | 0.00 | 8.87 | 7.55 | 1.32 |
| Las Vegas-Henderson-Paradise, NV | 3.02 | 3.02 | 0.00 | 0.00 | 15.26 | 14.08 | 1.18 |
| Los Angeles-Long Beach-Anaheim, CA | 3.16 | 2.62 | 0.54 | 0.03 | 15.46 | 14.54 | 0.92 |
| Louisville/Jefferson County, KY-IN | 2.88 | 2.88 | 0.00 | 0.00 | 12.56 | 12.56 | 0.00 |
| Memphis, TN-MS-AR | 1.69 | 1.69 | 0.00 | 0.00 | 12.46 | 12.46 | 0.00 |
| Miami-Fort Lauderdale-West Palm Beach, FL | 1.80 | 1.62 | 0.18 | 0.17 | 11.86 | 10.61 | 1.24 |
| Milwaukee-Waukesha-West Allis, WI | 3.43 | 3.35 | 0.07 | 0.00 | 13.70 | 12.64 | 1.06 |
| Minneapolis-St. Paul-Bloomington, MN-WI | 3.10 | 2.38 | 0.72 | 0.00 | 12.91 | 10.27 | 2.64 |
| Nashville-Davidson--Murfreesboro--Franklin, TN | 0.11 | 0.11 | 0.00 | 0.00 | 8.93 | 8.50 | 0.43 |
| New Orleans-Metairie, LA | 2.23 | 2.23 | 0.00 | 0.00 | 24.43 | 23.85 | 0.58 |
| New York-Newark-Jersey City, NY-NJ-PA | 11.68 | 4.01 | 7.67 | 1.24 | 28.05 | 26.87 | 1.18 |
| Oklahoma City, OK | 1.52 | 1.42 | 0.10 | 0.00 | 9.52 | 9.52 | 0.00 |
| Orlando-Kissimmee-Sanford, FL | 0.31 | 0.13 | 0.18 | 0.00 | 7.86 | 6.23 | 1.64 |

| MSA | | | | | | | |
|---|---|---|---|---|---|---|---|
| Philadelphia-Camden-Wilmington, PA-NJ-DE-MD | 2.98 | 2.02 | 0.97 | 0.61 | 20.27 | 19.32 | 0.94 |
| Phoenix-Mesa-Scottsdale, AZ | 1.92 | 1.42 | 0.50 | 0.00 | 11.88 | 10.26 | 1.62 |
| Pittsburgh, PA | 3.07 | 2.43 | 0.64 | 0.00 | 15.87 | 15.54 | 0.33 |
| Portland-Vancouver-Hillsboro, OR-WA | 3.85 | 2.93 | 0.92 | 0.00 | 18.44 | 15.83 | 2.60 |
| Providence-Warwick, RI-MA | 1.47 | 1.42 | 0.05 | 0.04 | 14.48 | 12.45 | 2.03 |
| Raleigh, NC | 1.35 | 1.35 | 0.00 | 0.00 | 12.13 | 11.46 | 0.67 |
| Richmond, VA | 0.74 | 0.74 | 0.00 | 0.00 | 8.50 | 7.87 | 0.64 |
| Riverside-San Bernardino-Ontario, CA | 1.45 | 1.43 | 0.02 | 0.01 | 8.76 | 8.43 | 0.33 |
| Rochester, NY | 2.17 | 2.17 | 0.00 | 0.00 | 12.27 | 11.40 | 0.87 |
| Sacramento--Roseville--Arden-Arcade, CA | 1.76 | 1.26 | 0.50 | 0.03 | 14.63 | 12.65 | 1.98 |
| St. Louis, MO-IL | 0.87 | 0.84 | 0.03 | 0.00 | 10.40 | 9.81 | 0.59 |
| Salt Lake City, UT | 1.39 | 0.27 | 1.12 | 0.33 | 14.75 | 14.36 | 0.39 |
| San Antonio-New Braunfels, TX | 2.84 | 2.84 | 0.00 | 0.00 | 9.59 | 8.94 | 0.65 |
| San Diego-Carlsbad, CA | 2.66 | 2.19 | 0.47 | 0.11 | 15.00 | 13.86 | 1.14 |
| San Francisco-Oakland-Hayward, CA | 5.92 | 3.44 | 2.48 | 0.84 | 21.88 | 20.00 | 1.88 |
| San Jose-Sunnyvale-Santa Clara, CA | 2.26 | 1.52 | 0.75 | 1.02 | 18.74 | 15.87 | 2.87 |
| Seattle-Tacoma-Bellevue, WA | 6.46 | 5.32 | 1.14 | 0.00 | 15.71 | 14.31 | 1.40 |
| Tampa-St. Petersburg-Clearwater, FL | 0.95 | 0.92 | 0.03 | 0.00 | 12.51 | 11.66 | 0.85 |
| Virginia Beach-Norfolk-Newport News, VA-NC | 1.08 | 0.72 | 0.36 | 0.00 | 9.54 | 9.22 | 0.33 |
| Washington-Arlington-Alexandria, DC-VA-MD-WV | 7.43 | 3.84 | 3.59 | 0.33 | 21.01 | 20.01 | 1.00 |

Source: Calculated by the authors from the 2017 NHTS.

Note: In order to isolate urban travel, the sample was limited to residents of urban areas and trips of 75 miles or less. MSA (Metropolitan statistical area) is defined by the U.S. Office of Management and Budget. For specific definitions or boundaries of each MSA, see: https://www.census.gov/programs-surveys/metro-micro/geographies/reference-maps.html. This table only covers MSA with a population of at least one million due to the data availability of the 2017 NHTS public version.

**Table A3 Average Trip Length (adjusted) by Mode and Income Class (in miles)**

| Mode of Transportation | Household Income | | | | | |
| --- | --- | --- | --- | --- | --- | --- |
| | Less than $25,000 | $25,000 to $49,999 | $50,000 to $74,999 | $75,000 to $99,999 | $100,000 and over | All |
| Total Auto | 6.1 | 7.1 | 7.8 | 8.4 | 8.3 | 7.7 |
|     SOV[a] | 6.1 | 6.9 | 8.0 | 8.6 | 8.9 | 7.9 |
|     HOV[b] | 6.1 | 7.2 | 7.7 | 8.2 | 7.8 | 7.4 |
| Total Transit | 6.7 | 8.2 | 9.2 | 9.7 | 12.8 | 9.1 |
|     Bus Transit[c] | 6.1 | 6.8 | 7.9 | 9.7 | 11.9 | 7.5 |
|     Subway/Light Rail/Streetcar[d] | 8.6 | 10.2 | 9.6 | 8.2 | 9.3 | 9.2 |
|     Commuter Rail[e] | 18.0 | 17.0 | 21.1 | 23.6 | 25.5 | 23.4 |
| Total Nonmotorized | 0.8 | 0.7 | 0.7 | 0.7 | 0.7 | 0.7 |
|     Walk | 0.7 | 0.6 | 0.5 | 0.5 | 0.6 | 0.6 |
|     Bicycle | 1.6 | 2.4 | 2.1 | 2.2 | 2.3 | 2.1 |
| School Bus | 3.7 | 3.8 | 4.6 | 3.8 | 4.4 | 4.1 |
| Taxicab | 4.8 | 4.6 | 3.7 | 7.8 | 6.8 | 5.7 |
| Other | 3.9 | 10.5 | 9.1 | 7.1 | 10.7 | 8.3 |
| All | 5.0 | 6.3 | 6.9 | 7.4 | 7.3 | 6.6 |

Source: Calculated by the authors from the 2017 NHTS.

Notes: In order to isolate urban travel, the sample was limited to residents of urban areas and trips of 75 miles or less.

a. SOV (single-occupancy vehicle) includes vehicles with driver and no passengers.
b. HOV (high-occupancy vehicle) includes vehicles with two or more occupants.
c. Bus transit includes public bus, commuter bus, private/shuttle bus and city-to-city bus.
d. Subway/Light Rail/Streetcar also include elevated rail.
e. Commuter rail includes suburban/regional rail systems and short-distance service provided by Amtrak.

**References**

Adler, P., & Florida, R. (2021). The rise of urban tech: how innovations for cities come from cities. *Regional Studies*, *55*(10-11), 1787-1800.

Aksoy, C. G., Barrero, J. M., Bloom, N., Davis, S. J., Dolls, M., & Zarate, P. (2022). *Working from home around the world* (No. w30446). National Bureau of Economic Research.

Baek, K., Lee, H., Chung, J. H., & Kim, J. (2021). Electric scooter sharing: How do people value it as a last-mile transportation mode?. *Transportation Research Part D: Transport and Environment*, *90*, 102642.

Bastian, A., Börjesson, M., & Eliasson, J. (2016). Explaining "peak car" with economic variables. *Transportation Research Part A: Policy and Practice*, *88*, 236-250.

Ben-Akiva, M. E. & Lerman, S. R. (1985). *Discrete choice analysis: theory and application to travel demand* (Vol. 9). MIT press.

Bhuiyan, J. (2018). Uber powered four billion rides in 2017. It wants to do more — and cheaper — in 2018. *Vox.* Link: https://www.vox.com/2018/1/5/16854714/uber-four-billion-rides-coo-barney-harford-2018-cut-costs-customer-service

Blumenberg, E., Brown, A., Ralph, K., Taylor, B. D., & Turley Voulgaris, C. (2019). A resurgence in urban living? Trends in residential location patterns of young and older adults since 2000. *Urban Geography*, *40*(9), 1375-1397.

Boarnet, M. G. (2011). A broader context for land use and travel behavior, and a research agenda. *Journal of the American Planning Association*, *77*(3), 197-213.

Brown, A. (2017). Car-less or car-free? Socioeconomic and mobility differences among zero-car households. *Transport Policy*, *60*, 152-159.

Brown, A. (2019). Redefining car access: Ride-hail travel and use in Los Angeles. *Journal of the American Planning Association*, *85*(2), 83-95.

Brown, A., Klein, N. J., Smart, M. J., & Howell, A. (2022). Buying Access One Trip at a Time: Lower-Income Households and Ride-Hail. *Journal of the American Planning Association*, 1-13.

Bureau of Transportation Statistics. (2020). U.S. Ton-Miles of Freight. Link: https://www.bts.gov/content/us-ton-miles-freight

Bureau of Transportation Statistics. (2022). National Transportation Altas Database. Link: https://www.bts.gov/ntad

Cao, X. J. (2012). The relationships between e-shopping and store shopping in the shopping process of search goods. *Transportation Research Part A: Policy and Practice*, *46*(7), 993-1002.
42

Stapleton, L., Sorrell, S., & Schwanen, T. (2017). Peak car and increasing rebound: A closer look at car travel trends in Great Britain. *Transportation Research Part D: Transport and Environment*, *53*, 217-233.

Su, R., McBride, E. C., & Goulias, K. G. (2021). Unveiling daily activity pattern differences between telecommuters and commuters using human mobility motifs and sequence analysis. *Transportation Research Part A: Policy and Practice*, *147*, 106-132.

Taylor, B. D., & Morris, E. A. (2015). Public transportation objectives and rider demographics: are transit's priorities poor public policy?. *Transportation*, *42*(2), 347-367.

United Nations. (2018). World Urbanization Prospects. Link: https://www.un.org/development/desa/publications/2018-revision-of-world-urbanization-prospects.html

U.S. Census Bureau. (2021). 2020 Census Data. Link: https://www.census.gov/programs-surveys/decennial-census/decade/2020/2020-census-results.html

Wang, K., & Akar, G. (2020). Will millennials drive less as the economy recovers: a postrecession analysis of automobile travel patterns. *Journal of Planning Education and Research*, 0739456X20911705.

Wang, X. (2019). Has the relationship between urban and suburban automobile travel changed across generations? Comparing Millennials and Generation Xers in the United States. *Transportation Research Part A: Policy and Practice*, *129*, 107-122.

Wang, K., & Wang, X. (2021). Generational differences in automobility: Comparing ' 'America's Millennials and Gen Xers using gradient boosting decision trees. *Cities*, *114*, 103204.

Wang, X. (2015). Assimilation of Southern California's Immigrants: A Double-Cohort Demographic Analysis on Commuting Modes. Working paper.

Wang, X. (2022). Impact of health on driving for ' 'America's older adults: A nationwide, longitudinal study. *Transport policy*, *120*, 69-79.

Wang, X., Lindsey, G., Schoner, J. E., & Harrison, A. (2016). Modeling Bike Share Station Activity: Effects of Nearby Businesses and Jobs on Trips to and from Stations. *Journal of Urban Planning and Development*, *142*(1), 04015001.

Westat. (2019). 2017 NHTS User Guide. *Federal Highway Adminstration, Washington D.C., U.S.A.*
46